\newcommand{\bra}{\langle}
\newcommand{\ket}{\rangle}
\newcommand{\ur}{{\bf r}}
\newcommand{\uR}{{\bf R}}

\newcommand{\uF}{{\bf F}}
\newcommand{\uB}{{\bf B}}
\newcommand{\uv}{{\bf v}}
\newcommand{\uy}{{\bf y}}
\newcommand{\dms}{\hat{\rho}}

\newcommand{\proj}{\hat{\mathbf{P}}}

\newcommand{\hks}{\hat{h}_{\text{KS}}}

\documentclass[%
aip,
jcp,%
amsmath,amssymb,
reprint,%
author-year,%
author-numerical,%
]{revtex4-2}

\usepackage{amsmath}
\usepackage{amsfonts}
\usepackage{amsthm}

\usepackage{natbib}
\usepackage{textcomp}
\usepackage[a4paper,margin=3cm]{geometry}
\usepackage{graphicx}
\usepackage{float}
\usepackage{hyperref}

\allowdisplaybreaks

\begin{document}

\title{Structure Optimization with Stochastic Density Functional Theory}
\author{Ming Chen}

\affiliation{Department of Chemistry, Purdue University, West
  Lafayette, Indiana 47907, USA}

\author{Roi Baer}

\affiliation{Fritz Haber Center of Molecular Dynamics and Institute of
  Chemistry, The Hebrew University of Jerusalem, Jerusalem, 91904
  Israel}

\author{Eran Rabani}

\affiliation{Department of Chemistry, University of California,
  Berkeley, California 94720, USA} \affiliation{Materials Sciences
  Division, Lawrence Berkeley National Laboratory, Berkeley,
  California 94720, USA } \affiliation{The Raymond and Beverly Sackler
  Center of Computational Molecular and Materials Science, Tel Aviv
  University, Tel Aviv 69978, Israel}

\begin{abstract}
Linear-scaling techniques for Kohn-Sham density functional theory
(KS-DFT) are essential to describe the ground state properties of
extended systems. Still, these techniques often rely on the locality
of the density matrix or on accurate embedding approaches, limiting
their applicability.  In contrast, stochastic density functional
theory (sDFT) achieves linear- and sub-linear-scaling by statistically
sampling the ground state density without relying on embedding or
imposing localization.  In return, ground state observables, such as
the forces on the nuclei, fluctuate in sDFT, making the optimization
of the nuclear structure a highly non-trivial problem.  In this work,
we combine the most recent noise-reduction schemes for sDFT with
stochastic optimization algorithms to perform structure optimization
within sDFT. We compare the performance of the stochastic gradient
descent (sGD) approach and its variations (stochastic gradient descent
with momentum (sGDM)) to stochastic optimization techniques that rely
on the Hessian, such as the stochastic Broyden-Fletcher-
Goldfarb-Shanno (sBFGS) algorithm. We further provide a detailed
assessment of the computational efficiency and its dependence on the
optimization parameters for each methods for determining the ground
state structure of bulk silicon with varying supercell dimensions.
\end{abstract}

\maketitle

\section{Introduction}
Kohn-Sham (KS) density functional theory
(DFT)~\cite{PhysRev.136.B864,PhysRev.140.A1133} is routinely used 
in determining the ground state properties of 
molecules~\cite{doi:10.1146/annurev.pc.46.100195.003413} and condensed
phases.\cite{doi:10.1002/wcms.1125} One of the key success of KS-DFT
is in determining the equilibrium
structures~\cite{doi:10.1021/acs.jctc.0c00549, math9222855} via
structural optimization routines, which often require many (tens)
single-point KS-DFT calculations.  Optimization procedures become 
prohibitively
expensive for large systems~\cite{doi:10.1021/acs.jctc.1c00237} due to 
the cubic ($O(N^3)$) scaling of conventional
KS-DFT.\cite{doi:10.1063/1.4947024}

Reducing the computational scaling of DFT is, thus, essential in order
to describe the equilibrium structures of extended systems.
Linear-scaling DFT has been highly fruitful in describing the
structure of extended bio-molecules and large-scale
materials,\cite{PhysRevB.47.9973,PhysRevB.48.14646,GOEDECKER1995261,PhysRevB.51.10157,PhysRevLett.76.3168,PhysRevB.58.12704,PhysRevLett.66.1438,PhysRevB.44.8454,PhysRevB.53.12713,PhysRevLett.79.3962,10.3389/fchem.2020.589910}
but relies on the assumption of ``near-sightedness'' of the one-body
reduced density
matrix,\cite{PhysRevB.47.9973,PhysRevLett.76.3168,PhysRevLett.79.3962}
$\rho(\ur, \ur')$.  This assumption works well for large band-gap
insulators,\cite{doi:10.1063/1.1839852,doi:10.1098/rsif.2013.0547,doi:10.1063/5.0005074}
but often fails for small band-gap materials, particularly, for
metals.\cite{doi:10.1063/1.4972007,doi:10.1063/1.4817001,MOHR201864}
Embedding methods can also achieve linear scaling by dividing the
extended system into small interacting
subsystems,\cite{PhysRevLett.66.1438,PhysRevB.44.8454,PhysRevB.53.12713}
however, designing an embedding method with accurate subsystem
interactions is still a challenging
problem.\cite{PhysRevB.53.12713,doi:10.1063/1.470549,doi:10.1021/ct9001784,
  doi:10.1021/cr500502v,doi:10.1063/1.3582913,doi:10.1063/1.3659293}

Recently developed stochastic density functional theory (sDFT)
achieves linear- and sub-linear
scaling,\cite{PhysRevLett.111.106402,doi:10.1063/1.4890651,doi:10.1063/1.5064472,
  doi:10.1063/1.5114984,doi:10.1063/5.0044163} without relying on the
sparsity of the density matrix nor on the design of the subsystem interaction in embedding schemes.
Instead, in sDFT, the electron density is represented using stochastic
orbitals (rather than Kohn-Sham (KS) orbitals) and linear scaling is
achieved by introducing a controlled statistical error in all ground
state observables, including the forces on the nuclei. However, this procedure 
poses
several challenges for determining the canonical equilibrium 
configurations using sDFT and for structural optimization. For
the former, the fluctuation-dissipation relations enable the utilization 
the noise in the nuclei forces to control the target
temperature.\cite{doi:10.1063/1.4984931,doi:10.1063/5.0004954}
However, using sDFT to determine the optimized structures requires
a low noise level on the nuclei forces.

Reducing the statistical noise in sDFT can be achieved by increasing
the number of stochastic orbitals.  However, decreasing the standard
deviation of the nuclei forces by one order of magnitude would require
an increase of two orders of magnitude in the number of stochastic
orbitals, significantly limiting the computational efficiency of sDFT.
Alternatively, we can control the noise level by using accurate
reference
schemes,\cite{doi:10.1063/1.4890651,doi:10.1063/1.5064472,doi:10.1063/1.5114984}
such as the most recent combination of real- and energy-space
fragmentation (``energy window embedded fragment stochastic density
functional theory'' (ew-efsDFT)).\cite{doi:10.1063/5.0044163} Here,
deterministic DFT (dDFT) determines the ground state density matrix
for each fragment, while sDFT is used to evaluate the corrections to
the fragment density.

Unfortunately, noise reduction schemes can not entirely eliminate the
noise on the nuclei forces, therefore, special structural optimization
techniques are still required for fluctuating forces.  A natural
choice is to use stochastic optimization methods often found in
machine learning and numerous optimization
algorithms.\cite{10.1214/aoms/1177729586,Rumelhart:1986we,doi:10.1137/0330046,JMLR:v12:duchi11a,tieleman2012lecture,adam,pmlr-v2-schraudolph07a,NIPS2013-ac1dd209}
Some scenarios assume that it is possible to access the deterministic
forces (offline methods)~\cite{NIPS2013-ac1dd209} while others rely
solely on the available noisy forces (online
methods).\cite{10.1214/aoms/1177729586,Rumelhart:1986we,doi:10.1137/0330046,JMLR:v12:duchi11a,tieleman2012lecture,adam,pmlr-v2-schraudolph07a}

In this work we use sDFT to obtain the forces on the nuclei and assess
the accuracy and performance of several online structural optimization
techniques in determining the optimized structures of bulk silicon
supercells.  We focus on several stochastic optimization methods
including the stochastic gradient descent
(sGD),\cite{10.1214/aoms/1177729586} the stochastic gradient descent
with momentum (sGDM),\cite{Rumelhart:1986we} and the stochastic
Broyden-Fletcher-Goldfarb-Shanno (sBFGS).\cite{pmlr-v2-schraudolph07a}
The manuscript is organized as follows: sDFT and various noise
reduction techniques are introduced in Sections~\ref{sec:sDFT} and
\ref{sec:noise}, respectively. Section~\ref{sec:smin} summarizes
several stochastic optimization methods used in this work.
Section~\ref{sec:rd} compares of the different optimization schemes
for bulk silicon supercells and provides a detailed analysis of the
optimization performance as a function of size, the number of
stochastic orbitals, damping parameter, and optimization step size.
Finally, in Section~\ref{sec:con} we summarize and discuss the
findings.

\section{Stochastic Density Functional Theory}
\label{sec:sDFT}
In KS-DFT, the one-particle KS-Hamiltonian is given by:
\begin{equation}
\hat{h}_{\text{KS}}[\rho]=\hat{t}+\hat{v}_{\text{loc}}+\hat{v}_{\text{nl}}+
\hat{v}_{\text{H}}[\rho]+\hat{v}_{\text{xc}}[\rho]\;\;,
\end{equation}
where $\hat{t}$, $\hat{v}_{\text{loc}}$, $\hat{v}_{\text{nl}}$,
$\hat{v}_{\text{H}}$, and $\hat{v}_{\text{xc}}$ are kinetic operator,
local pseudopotential operator, non-local pseudopotential operator,
Hartree operator and exchange-correlation operator. The KS Hamiltonian
depends on electron density $\rho$, defined as (for clarity, we ignore
spin polarization):
\begin{equation}
\rho(\mathbf{r})=2\sum_{i=1}^{N_{\text{occ}}}|\phi_i(\mathbf{r})|^2 
\;\;,
\end{equation}
where $N_{\text{occ}}$ is the number of occupied orbitals and
$\phi_i(\mathbf{r})$ is the $i$'th KS orbital solved by diagonalizing
$\hat{h}_{\text{KS}}$.

In sDFT, the electron density $\rho(\mathbf{r})$ is represented as an
average over stochastic orbitals,\cite{doi:10.1146/annurev-physchem-090519-045916}
$\{\chi(\mathbf{r})\}$, i.e.~\cite{PhysRevLett.111.106402}
\begin{equation}
\rho(\ur)=2\left\langle\langle\chi|\hat{\rho}\delta
(\hat{\mathbf{r}}-\mathbf{r})|\chi\rangle\right\rangle_{\chi}
=2\left\langle|\xi(\mathbf{r})|^2\right\rangle_\chi\;\;,
\label{eq:rho-sdft}
\end{equation}
where $\langle\cdots\rangle_\chi$ denotes averaging over all
realizations of $\chi$. In practice, only a finite number ($N_\chi$)
of stochastic orbitals are used and
$\langle\cdots\rangle_\chi=\frac{1}{N_\chi}\sum_{j}^{N_\chi}
|\xi_j(\mathbf{r})|^2 $.  In the above equation,
$\hat{\rho}=\sum_{i=1}^{N_{\text{occ}}}|\phi_i\rangle\langle\phi_i|$
is the one-body density matrix, approximated by $\hat{\rho} \approx
f(\hat{h}_{\text{KS}},\mu,\beta)$ in sDFT, where $f(x)$ is the
Fermi-Dirac distribution function parameterized by the chemical
potential $\mu$ and inverse temperature $\beta$.

Using a real-space representation, for example, a stochastic orbital
$\chi(\mathbf{r})$ takes random values $\pm 1/\sqrt{\Delta V}$ at each
grid point, where $\Delta V$ is the volume element of the real space
grid. Projecting $\chi(\mathbf{r})$ onto the occupied space to
generate $\xi(\mathbf{r})$ in Eq.~\eqref{eq:rho-sdft} is done by
expanding the density matrix in a Chebyshev
series:\cite{Kosloff1988,Kosloff1994,PhysRevLett.79.3962}
\begin{equation}
\sqrt{f(\hat{h}_{\text{KS}},\mu,\beta)}= \sum_{n=0}^{N_{\text{c}}}\alpha_n(\mu,\beta)
T_n(\hat{h}_{\text{KS}}),
\label{eq:sdft-cheby}
\end{equation}
where $N_{\mathrm{c}}$ is the length of the polynomial,
$\alpha_n(\mu,\beta)$ and $T_n(\hat{h}_{\text{KS}})$ are the expansion
coefficient and the Chebyshev polynomial of order $n$,
respectively. Evaluating the Chebyshev series in the above requires
iteratively applying $\hat{h}_{\text{KS}}$ on a stochastic orbital,
which is achieved with a linear scaling computational cost ($O(N_g)$,
where $N_g$ is the size of the grid) for the real space grid
representation.

In addition to describing the electron density using stochastic
orbitals, sDFT is capable of evaluating other one-body observable:
\begin{equation}
O=\mathrm{Tr}(\hat{\rho}\hat{O})=\left\langle\langle\xi|\hat{O}
|\xi\rangle\right\rangle_\chi
\label{eq:1b-ob-sdft}
\end{equation}
Since only a finite number of stochastic orbitals are used, the
expectation value of $\hat{O}$ calculated from sDFT fluctuate. For
many observables such as the electron density, the density of states,
and the forces on the nuclei, the stochastic errors do not depend on
the system size, and thus, only a small number of $N_\chi$ (relative
to the total number of occupied KS orbitals) is required as the system
size increases.

\section{Noise Reduction Schemes in Stochastic Density Functional Theory}
\label{sec:noise}
Over the last several years, we have developed several schemes to
reduce the noise in
sDFT.\cite{doi:10.1063/1.4890651,doi:10.1063/1.5064472,doi:10.1063/5.0044163}.
Currently, the most efficient of these~\cite{doi:10.1063/5.0044163} is
based on two simple procedures: overlapped fragments and energy
windowing.  In the first procedure, the system is divided into
overlapping fragments~\cite{doi:10.1063/1.5064472} and the total
electron density matrix is described by a sum of a reference density
matrix and a correction term:
\begin{equation}
\dms = \dms_{\mathrm{ref}}+\bra|\xi\ket\bra\xi|\ket_\chi-\sum_f|\xi^f\ket\bra\xi^f|\;\;.
\label{eq:dm-oefsDFT}
\end{equation}
In the above equation, the reference density is given by
$\dms_{\mathrm{ref}}=\sum_f
\sum_{i=1}^{N^f_{\mathrm{occ}}}|\varphi_i^f\ket \bra\varphi_i^f|$,
where $|\varphi_i^f\rangle$ is a KS orbital of fragment $f$ and
$N^f_{\mathrm{occ}}$ is the number of occupied orbitals of the $f$'th
fragment. Stochastic orbitals are used to sample the difference
between the system density matrix $\dms$ and the reference density
matrix $\dms_{\mathrm{ref}}$,
i.e. $|\xi^f\ket=\sum_{i=1}^{N^f_{\mathrm{occ}}}|\varphi_i^f\ket
\bra\varphi_i^f|\chi\ket$. Statistical errors in the second and the
third terms on the right hand side of Eq.~\eqref{eq:dm-oefsDFT} cancel
to a great extent as long as the reference density matrix
($\dms_{\mathrm{ref}}$) is closed to full one ($\dms$). This scheme
reduces the noise on the nuclei forces and other one-body observables
by a factor of $\approx 4-5$, as demonstrated for semiconductor
materials.\cite{doi:10.1063/1.5064472}

The energy windowing is the second procedure, which leads to further
noise reduction. It involves dividing the occupied space into energy
windows~\cite{doi:10.1063/1.5114984,doi:10.1063/5.0044163} and then
representing the identity operator $\mathbf{I}$ as a sum of projectors
onto these energy windows, $\mathbf{I}=\sum_{i=1}^{N_w}\proj_i$, where
$\proj_0=f(\hks,\varepsilon_0,\beta)$,
$\proj_i=f(\hks,\varepsilon_i,\beta)- f(\hks,\varepsilon_{i-1},\beta)$
for $i=1,\cdots,N_w-1$, $\proj_{N_w}=
\mathbf{I}-f(\hks,\varepsilon_{N_w-1},\beta)$, and $N_w$ is the number
of energy windows.  Using this representation for the identity
operator, the system density matrix can be written as:
\begin{equation}
\dms= \dms_{\mathrm{ref}}+\sum_{i=1}^{N_w}\bra|\xi_i\ket\bra\xi_i|\ket_\chi-
\sum_f\sum_{i=1}^{N_w}\bra|\xi_i^f\ket\bra\xi_i^f|\ket_\chi\;\;,
\label{eq:dm-ew-efsDFT}
\end{equation}
where $|\xi_i\ket=\sqrt{\dms\proj_i}|\chi\ket$ and
$|\xi_i^f\ket=\sum_{i=1}^{N^f_{\mathrm{occ}}}|\varphi_i^f\ket
\bra\varphi_i^f|\sqrt{\proj_i}|\chi\ket$. This scheme offers further
reduction of the noise on the nuclei forces, by roughly
$1/\sqrt(N_w)$ for $N_w \le 40$.\cite{doi:10.1063/5.0044163}

\section{Stochastic Minimization}
\label{sec:smin}
For a finite number of stochastic orbitals, it is impossible to
completely eliminate the noise on the nuclei forces.  Thus, structural
optimization of extend system relies on stochastic electronic methods
and poses challenges for optimization schemes due to the fluctuating
nature of the forces on the nuclei. Before discussing the performance
of the different stochastic optimization schemes considered in this
work, we briefly outline each approach.

The stochastic gradient descent with momentum (sGDM) and its
variations~\cite{adam,tieleman2012lecture} have been widely used in
optimizing neural networks in machine learning.\cite{lan2020first} In
sGDM method, the positions ($\uR_{n}$) and descent direction
($\uv_{n}$) of all atoms in optimization step $n$, is updated
according to:
\begin{subequations}
\begin{align}
\uv_{n+1} & =\gamma \uv(n)+ \uF(\uR_n,\delta_n)\Delta x_n, 
\label{eq:msgd1} \\
\uR_{n+1} & = \uR_n+\uv_{n+1}\;\;\ldotp
\label{eq:msgd2}
\end{align}
\end{subequations}
In the above, $\uF(\uR_n,\delta_n)$ is the force on the nuclei in step
$n$ and $\delta_n$ is the random seed used to obtain the force on the
nuclei for step $n$.  Thus, at each optimization step we change the
random seed to generate the stochastic orbitals and hence the nuclei
forces obtained from sDFT.  $\Delta x_n$ defines the step size and
$0\leq \gamma<1$ controls the degree of ``friction''.  For $\gamma=0$,
sGDM reduces to the stochastic gradient descent (sGD). sGDM is
guaranteed to converge to a local minimum as long as (a)
$\sum_{n=0}^{\infty}\Delta x_n=\infty$ and (b)
$\sum_{n=0}^{\infty}\Delta x_n^2<\infty$.
A typical choice is $\Delta x_n\propto 1/n$.  These conditions can be
satisfied by decreasing the step size during the optimization
trajectory.\cite{10.1214/aoms/1177729586,9264458}
$\sum_{n=0}^{\infty}\Delta x_n^2<\infty$ ensures $\Delta x_n$ decay
fast enough so that the noise of $\uF(\uR_n,\delta_n)$ sufficiently
decays by scaling the noise with $\Delta x_n$.
$\sum_{n=0}^{\infty}\Delta x_n=\infty$ prevents $\Delta x_n$ from
decaying too fast such that the optimization stops before the system
reaches a local minimum. Although it has been proved that a stochastic
minimization is guaranteed to converge to a local minimum regardless
of the choice of $\Delta x_n$ as long as condition (a) and (b) are
satisfied,\cite{10.1214/aoms/1177729586} the convergence rate becomes
rather slow.\cite{doi:10.1137/070704277}

The behavior of the optimization trajectories in sGDM for a fixed
$\Delta x_n$ is worth mentioning.  Typical optimization trajectories
can be divided into two stages.  In the first (descent) stage, the
average force $\bra\uF(\uR_n,\delta_n)\ket$ is much larger then its
fluctuations. In this stage, the efficiency of sGDM is similar to that
of the corresponding deterministic gradient descent approach and the
role of $\gamma$ is mainly to average the forces on the nuclei over
previous steps. This averaging also helps overcome instabilities
associated with ill-conditioned Hessians, which often results in a
trajectories that do not follow the descent direction for
$\gamma=0$. On the other hand, if $\gamma$ is too large $\uv_n$ has a
long-term memory so that $\uv_n$ can not represent the force on the
nuclei at the current configuration. Section~\ref{sec:rd} will discuss
the optimal choice of $\gamma$. As $\bra\uF(\uR_n,\delta_n)\ket$
decreases during the optimization, the magnitude of
$\bra\uF(\uR_n,\delta_n)\ket$ becomes comparable to the fluctuation of
$\uF(\uR_n,\delta_n)$, and the optimization enters the second
(averaging) stage where a cluster of configurations forms about a
local minimum. The cluster size can be controlled by the noise level,
which, thus, determines the accuracy of the relaxed structures.

While sGDM can significantly improve the efficiency and accuracy of
stochastic optimization compared to sGD, more efficient schemes also rely 
on information from the Hessian $\mathbf{H}$, such as the
stochastic Broyden-Fletcher-Goldfarb-Shanno (sBFGS).  In sBFGS, the
equations of updating $\uR_n$ are
\begin{subequations}
\begin{align}
\uv_n & = \uB_n\uF(\uR_n;\delta_n)\frac{\Delta x_n}{c}\;\;, 
\label{eq:SBFGS1} \\
\uR_{n+1} & = \uR_n+\uv_n \;\;,
\label{eq:SBFGS2} \\
\uy_n & = \uF(\uR_n;\delta_n)-\uF(\uR_{n+1};\delta_n) + \lambda \uv_n\;\;,
\label{eq:SBFGS3} \\
s_n & = (\uv_n^\top\uy_n)^{-1} \;\;,
\label{eq:SBFGS4} \\
\uB_{n+1} & = (\mathbf{I}-s_n\uv_n\uy_n^\top)\uB_n(\mathbf{I}-s_n\uy_n\uv_n^\top)+ \nonumber \\
& \quad\quad cs_n\uv_n\uv_n^\top \;\;\ldotp
\label{eq:SBFGS5}
\end{align}
\end{subequations}
In the above, $\mathbf{I}$ is the identity matrix, as before,
$\uF(\uR_n,\delta_n)$ is the force on the nuclei at configuration
$\uR_n$ with random number seed $\delta_n$ while
$\uF(\uR_{n+1},\delta_n)$ is the nuclei force at configuration
$\uR_{n+1}$ with the same random number seed $\delta_n$. Therefore,
two sDFT calculations (that can be performed simultaneously) are
required in each optimization iteration in sBFGS.  The other two
controlled parameters are $0<c\leq 1$, which was shown empirically to
improve the performance of sBFGS~\cite{pmlr-v2-schraudolph07a} and
$\lambda\leq 0$, which guarantees that $\mathbf{B}$ converges to
$(\mathbf{H}+\lambda\mathbf{I})^{-1}$ rather than to $\mathbf{H}^{-1}$
and thus, ensures that $\mathbf{B}$ is positive
definite.\cite{pmlr-v2-schraudolph07a} In the applications reported
below, in order to compare the optimization efficiencies for the same
step size in sGDM and sBFGS, we scale
$\mathbf{B}_n\uF(\uR_{n},\delta_n)$ by
$\|\uF(\uR_{n},\delta_n)\|/\|\mathbf{B}_n\uF(\uR_{n},\delta_n)\|$, so
that the magnitude of $\mathbf{B}_n\uF(\uR_{n},\delta_n)$ is the same
as $\uF(\uR_{n},\delta_n)$. We want to emphasize that preconditioning
force with Hessian significantly improve the sampling efficiency of
Langevin dynamics.\cite{doi:10.1063/5.0004954}

For the case of $c=1$ and $\lambda=0$, the above algorithm reduces to
the deterministic BFGS.  In deterministic BFGS, $\Delta x$ is usually
determined by a line search algorithm to ensure a sufficient descent
of the energy along the direction of $\uB_n\uF(\uR_n)$.  However, in
sDFT, the fluctuation of total energy increases with the system size
and are therefore challenging to evaluate.  Therefore, all sBFGS
calculations reported below did not use a line search for determining
$\Delta x$.

\section{Results and Discussion}
\label{sec:rd}
To test the accuracy and convergence of the different stochastic
optimization algorithms, we studied the optimization trajectories of
bulk silicon. We compared the stochastic results to deterministic
calculations for two system sizes, Si$_{216}$ and Si$_{512}$,
corresponding to a supercell of $3\times 3\times 3$ and $4\times
4\times 4$ unit cells, respectively. All DFT calculations (stochastic
and deterministic) were performed using a plane wave/real space grid
representation, within the local density approximation (LDA)
functional. We wish to point out that bulk silicon has an LDA band gap
of~\cite{doi:10.1063/1.2796168} $\approx$0.6 eV, which is challenging
for linear-scaling DFT methods, and serves as a challenging test for
sDFT. We used 30/60 Ryd for the wavefunction and the density cutoffs,
respectively. The Troullier-Martins norm-conserving
pseudopotentials~\cite{PhysRevB.43.1993} in the Kleinman-Bylander
form~\cite{PhysRevLett.48.1425} were used, and a real-space
implementation of the non-local pseudopotential was adopted to reduce
computational cost.\cite{PhysRevB.44.13063} To converge the ground
state properties, we took the value of $\beta\approx 600$ Ha$^{-1}$ in
the Chebyshev expansion of the density matrix (cf.,
Eq.~\eqref{eq:sdft-cheby}). $80$ stochastic orbitals were used in all
o-efsDFT/ew-efsDFT calculations and 41 energy windows were used for
ew-efsDFT calculations, unless otherwise noted. $2\times 2\times 2$
supercells were used as overlapped fragments in o-efsDFT and ew-efsDFT
while a $1\times 1\times 1$ unit cell was selected as a non-overlapped
region in each
fragment.\cite{doi:10.1063/1.4890651,doi:10.1063/1.5064472,doi:10.1063/1.5114984,doi:10.1063/5.0044163}

\begin{figure}[t]
\centering \includegraphics[width=7cm]{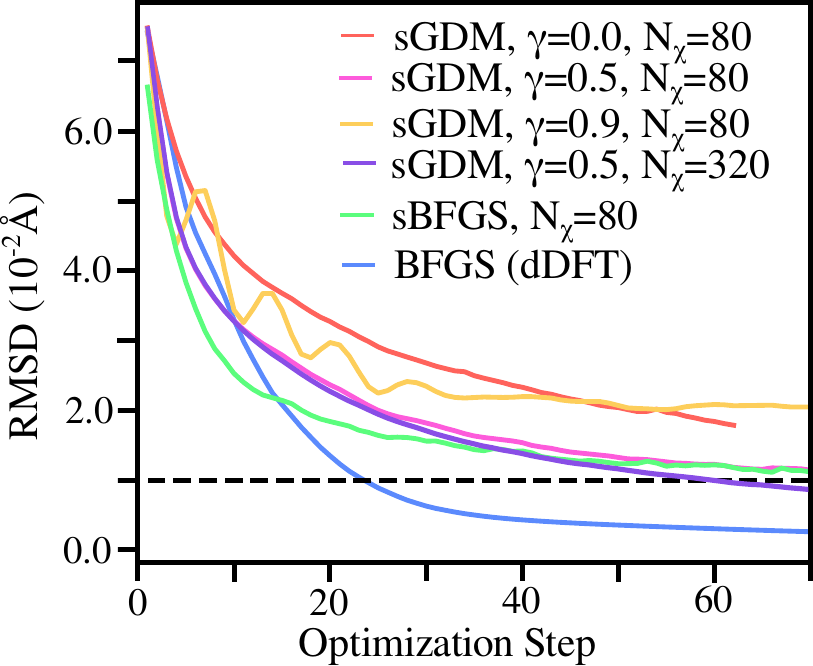}
\caption{Comparison of the RMSD along an optimization trajectory for
  $\mbox{Si}_{216}$ using ew-efsDFT with $N_\chi=80$ for various
  stochastic optimization techniques.  The blue and green curves
  depict the deterministic BFGS results (reference calculation) and
  sBFGS, respectively.  sGDM with $\gamma=0$, sGDM($\gamma=0.5$), and
  $\gamma=0.9$ are shown in red, magenta, and orange,
  respectively. For $\gamma=0.5$ we also show results with
  $N_\chi=320$ stochastic orbitals (purple curve).}
\label{fig:diff-opt}
\end{figure}

Typical optimization trajectories for Si$_{216}$ are shown in
Fig.~\ref{fig:diff-opt}, where we plot the root-mean-square-distance
(RMSD) between the current structure and the equilibrium structure
(obtained from using deterministic DFT) as a function of the
optimization step. The stochastic results are compared to a
deterministic DFT optimization using the BFGS method, which converges
monotonically to the optimized structure and requires $\approx 20-30$
steps to reach chemical accuracy (RMSD$\le 0.01\AA$ - dashed
horizontal line).  Comparing the different stochastic optimization
algorithms, we find, as expected, that the sGD (sGDM with $\gamma=0$)
requires over $60$ steps to reach and RMSD of $\approx 0.02\AA$. The
slow descent rate of sGD results from the ill-conditioned Hessian
matrix of a large solid state system. The convergence is much faster
when $\gamma$ takes a finite value (sGDM with $\gamma = 0.5$),
however, as $\gamma \rightarrow 1$ (sGDM with $\gamma = 0.9$) the
optimization trajectory does not follow the descent direction and the
RMSD oscillates with the optimization step. Furthermore, the RMSD of
the optimized structure is rather large ($\approx 0.02\AA$) compared
to the other stochastic approaches shown, with the same level of
statistical noise.

In order to better understand the role of the noise and on the
optimized structure, we assumed that the force covariance matrix,
$\mathbf{\Sigma}$, is diagonal. In this case, Eqs~.\eqref{eq:msgd1}
and \eqref{eq:msgd2} are simply a discretized version of the following
Langevin equation:
\begin{equation}
m\ddot{\uR}=-m\eta \dot{\uR}+\bra\uF\ket+
\sqrt{2k_{\mathrm{B}}T_{\mathrm{eff}}m\eta}~\mathbf{W}(t),
\label{eq:lg}
\end{equation}
where the mass $m=\Delta x$, the friction $\eta=(1-\gamma)/\Delta x$, and
$\mathbf{W}(t)$ is white noise (see supplementary information for more
information). In the above equation, $k_{\mathrm{B}}$ is the Boltzmann
constant and $T_{\mathrm{eff}}$ is an effective temperature:
\begin{equation}
T_{\mathrm{eff}}=\frac{\Delta x\sigma^2}{2k_{\mathrm{B}}(1-\gamma)}\;\;\ldotp
\label{eq:teff}
\end{equation}
The invariant probability distribution of Eq.~\eqref{eq:lg} is the
Boltzmann distribution:
\begin{equation}
P(\uR) = \frac{1}{Q}e^{-\frac{V(\uR)}{k_{\mathrm{B}}T_{\mathrm{eff}}}}
\label{eq:prob-inv}
\end{equation}
where $Q$ is the partition function and $V(\uR)$ is the potential
energy function that depends on the nuclei positions.  From
Eqs.~\eqref{eq:teff} and \eqref{eq:prob-inv} we can conclude (a) The
effective temperature is linearly dependent on the step size, $\Delta
x$ and (b) The effective temperature is inversely proportional to
$\gamma$. We note in passing that there are different variants of sGD
with adaptive step size like RMSProp~\cite{tieleman2012lecture} and
Adam~\cite{adam}, which are usually important in the averaging stage.

The optimization results of sGDM with $N_\chi=80$ and $N_\chi=320$
suggest that reducing noise in sDFT is not helpful in the early stage
of the optimization. The magnitude of the force, $\langle
\uF\rangle$, is much larger then $\sigma$ in the descent stage and
thus, increasing $\sigma$ does not change the descent
direction significantly. Therefore, one can use a small number of stochastic
orbitals at the descent stage and increase the number of stochastic
orbitals as the optimization progresses to the averaging
stage. Comparing the results in Fig.~\ref{fig:diff-opt} at early and
later stages of the optimization for $N_{\chi}=80$ and $N_{\chi}=320$
stochastic orbitals, clearly show the advantage of using more
stochastic orbitals at the averaging stage. Increasing the number
of stochastic orbitals along the optimization trajectory (``on the
fly'') is analogous to increasing the batch sizes in sGD optimizations
for machine learning.\cite{pmlr-v54-de17a}

\begin{figure*}[ht]
\centering \includegraphics[width=14cm]{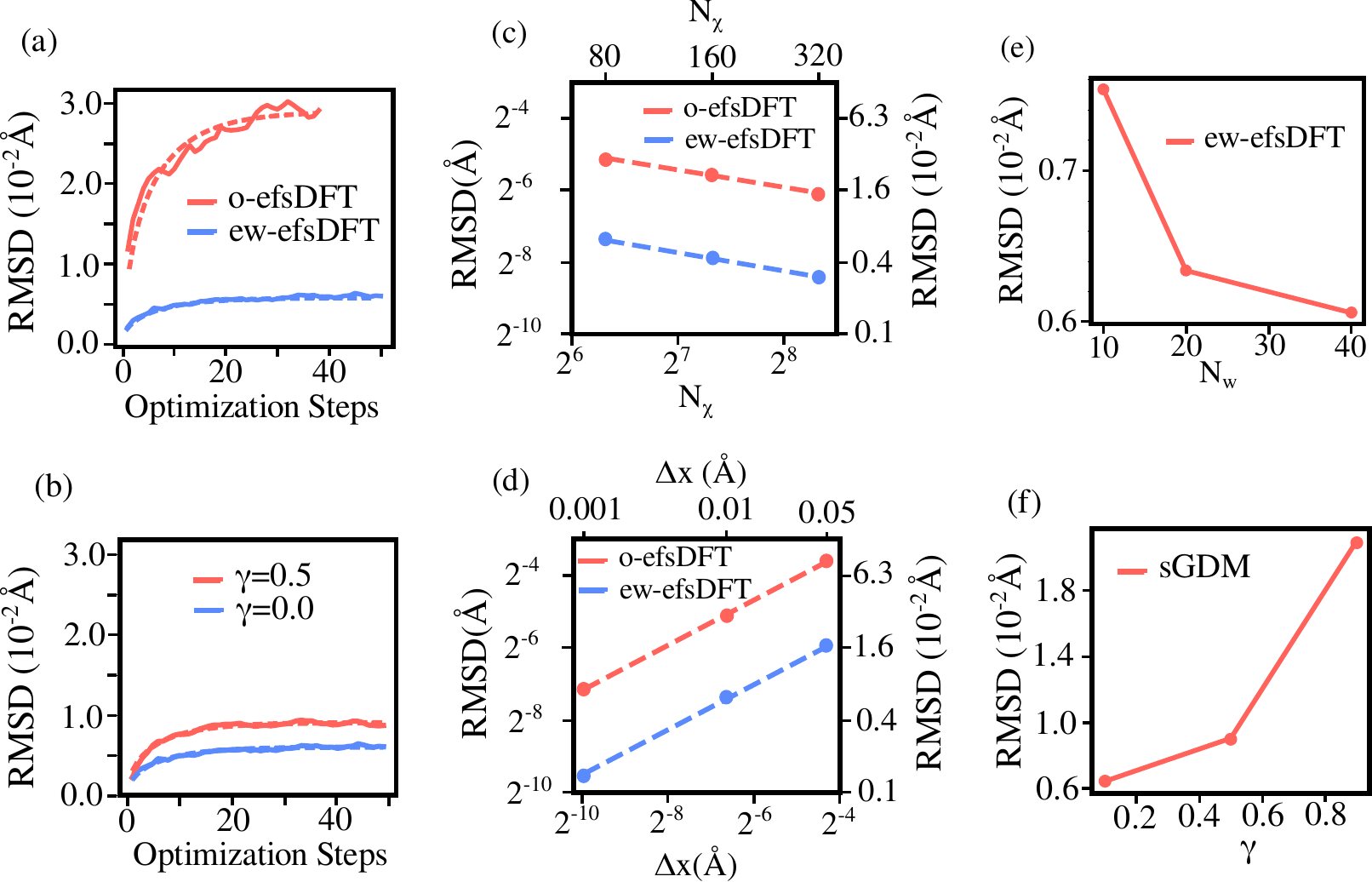}
\caption{(a) Reverse optimizations trajectories using sGD for o-efsDFT
  (red solid line) and ew-efsDFT (blue solid line). Dashed lines are
  fits to Eq.~\eqref{eq:ou-var}. (b) Reverse optimization trajectories
  comparing sGD (solid blue line) and sGDM (solid red line) with
  $\gamma=0.5$.  Dashed lines are fits to Eq.~\eqref{eq:ou-var}.  (c)
  A log-log plot of the variance in the nuclei positions, RMSD vs. the
  number of stochastic orbitals, $N_\chi$, using with o-efsDFT (red
  symbols) and ew-efsDFT (blue symbols). Power law fits
  ($\mathrm{RMSD} \propto N_\chi^{-\alpha}$) are shown by the dashed
  lines, with exponents given by $\alpha=0.48$ and $\alpha=0.51$,
  respectively. (d) A log-log plot of the variance in the nuclei
  positions, RMSD vs. the optimization step size, $\Delta x$, using
  with o-efsDFT (red symbols) and ew-efsDFT (blue symbols). Power law
  fits ($\mathrm{RMSD} \propto \Delta x^{\eta}$) are shown by the
  dashed lines, with exponents given by $\eta=0.62$ for both
  calculations. (e) The optimized RMSD vs. the number of windows used
  in ew-efsDFT, $N_w$. (f) The RMSD of optimized structure
  vs. $\gamma$ using ew-efsDFT.}
\label{fig:param}
\end{figure*}

To better understand the behavior of the different optimization
approaches in the averaging stage, we initiated optimization
trajectories from the equilibrium structure obtained by deterministic
DFT and analyzed the behavior of the RMSD for different noise levels,
friction, and step sizes. The results are summarized in
Fig.~\ref{fig:param}. Two such optimization trajectories are shown in
Fig.~\ref{fig:param}(a) for o-efsDFT (red curve) and ew-efsDFT (blue
curve) using sGDM with $\gamma=0$ and $\Delta x=0.01$. The RMSD
increases from its optimal value of $0$, approaching a plateau at long
times, resulting from the noisy forces in both sDFT methods. Since the
fluctuations of the forces on the nuclei in o-efsDFT are larger than
those in ew-efsDFT (as a result of using energy
windowing,\cite{doi:10.1063/5.0044163}) the plateau value of the RMSD
is significantly larger, and the approach to the plateau is slower in
the former.

Both optimization trajectories fluctuate about the optimized
structure, and thus, the forces on the nuclei can be approximated
by Hooke's law. In this limit, a reversed optimization trajectory for
sGDM is equivalent to a discretized version of an Ornstein-Uhlenbeck
(O-U) process (see SI for more information) for a small $\Delta x$:
\begin{equation}
\dot{x} = -kx+\sigma W(t).
\end{equation}
In the above, $k$ is the force constant and $W$ is the random white
noise. The variance of $x$ at time $t$ for the above is given
by:\cite{varadhan2007stochastic}
\begin{equation}
\mathrm{Var}(x(t)) \approx \frac{\Delta x\sigma^2}{2k(1-\gamma)}\left(1-e^{-2kt/(1-\gamma)}\right)
\label{eq:ou-var}
\end{equation}
Inspired by Eq.~\eqref{eq:ou-var}, we fitted the RMSD curves shown in
Fig.~\ref{fig:param} panels (a) and (b) to Eq.~\eqref{eq:ou-var}
(dashed curves), with $\sigma$ and $k$ used as free parameters. The
fits seem to describe the numerical data quite accurately. The above
expression suggests that increasing $\gamma$ should result in a larger
RMSD plateau and fast converges, which is indeed the numerical case
shown in Fig.~\ref{fig:param}(b), reconfirming Eq.~\eqref{eq:teff}.

\begin{figure}[t]
\centering \includegraphics[width=7cm]{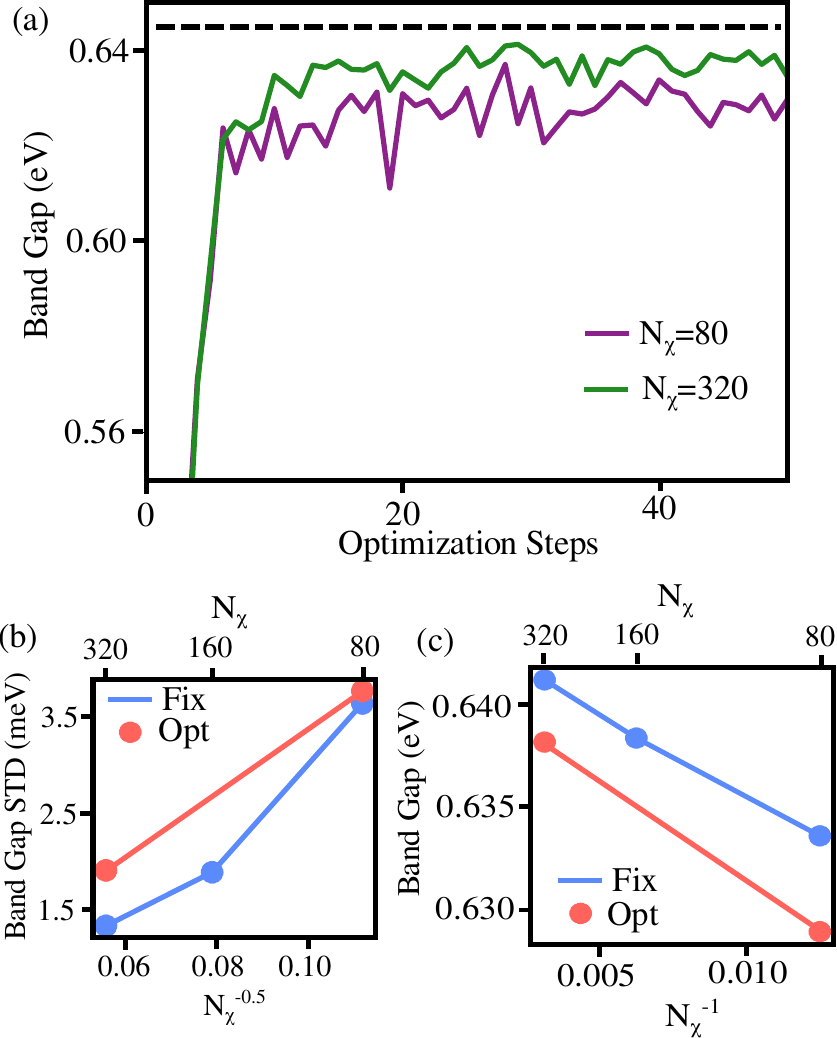}
\caption{(a) The fundamental band gap along an optimization trajectory
  using ew-efsDFT and sGDM with $\gamma = 0.4$. Purple and green curves
  correspond to $N_\chi=80$ and $N_\chi=320$, respectively. The black
  dashed line is the band gap calculated by deterministic DFT for the
  optimized structure.  (b) and (c) show the standard deviation of
  the band gap and the average band gap calculated during the averaging
  stage of the optimization (red symbols) and for the equilibrium
  structure (blue symbols) vs. the inverse number of stochastic
  orbitals.}
\label{fig:gap}
\end{figure}

\begin{figure}[t]
\centering \includegraphics[width=7cm]{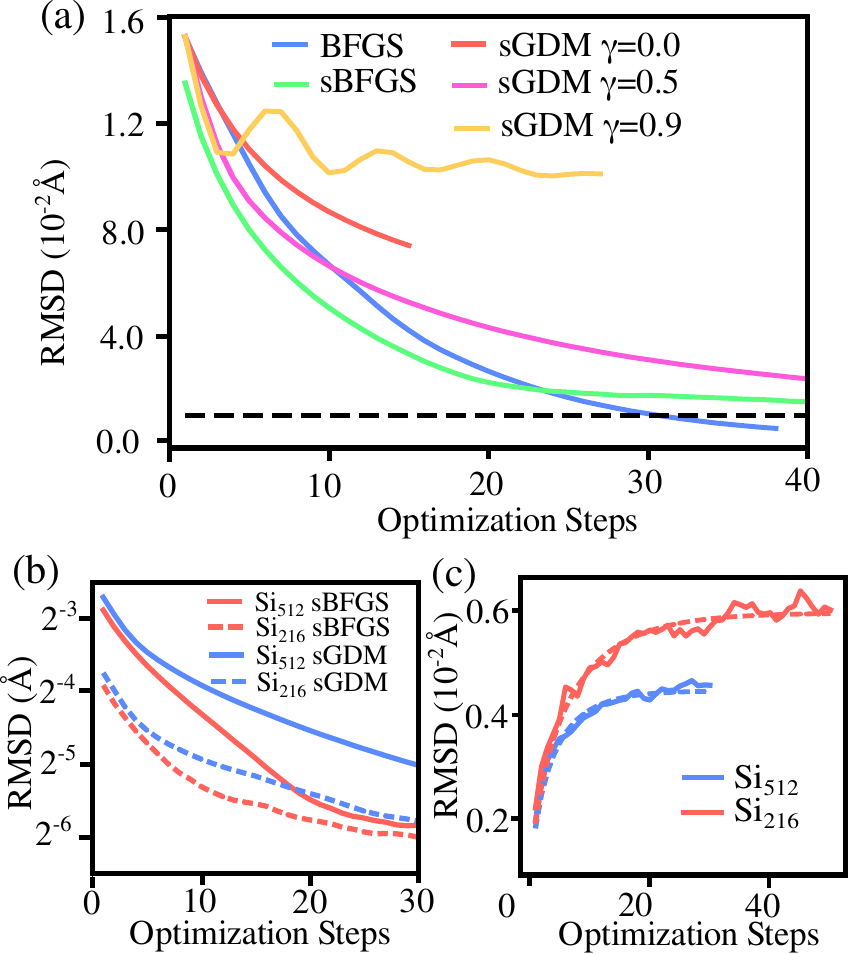}
\caption{(a) Same as Fig.~\ref{fig:diff-opt} for
  $\mbox{Si}_{512}$. (b) Comparison of RMSD along an optimization
  trajectory for Si$_{216}$ (dashed lines) and Si$_{512}$ (solid
  lines). (c) Same as panel (b) for a reverse optimization. }
\label{fig:512}
\end{figure}

In Fig.~\ref{fig:param} panels (c) and (d) we show the variance in the
position of the nuclei (RMSD) as a function of the number of
stochastic orbitals and the step size, respectively. We find that
$\mathrm{RMSD} \propto N_\chi^{-0.5}$ and $\mathrm{RMSD}\propto \Delta
x^{0.62}$, in close agreement with the expected statistical
values. Panels (c) and (d) of Fig.~\ref{fig:param} also show that
ew-efsDFT leads to a much better-optimized structure compared to
o-efsDFT regardless of parameters used in optimizations. However,
increasing $N_w$ above $20$ in ew-efsDFT only marginally improves the
results, as shown in Fig.~\ref{fig:param}(e).  Finally, in
Fig.(\ref{fig:param})(f) we show the RMSD as a function of $\gamma$,
which is a non-linear function of $\gamma$, consistent with
Eq.~\eqref{eq:ou-var}. The numerical ratios of RMSD with $\gamma=0.1$,
$0.5$ and $0.9$ are $1:1.41:3.28$ in good agreement with the predicted
values based on Eq.~\eqref{eq:ou-var} ($1:1.41:3.0$).

In Fig.~\ref{fig:gap}(a) we plot the fundamental band gap along an
optimization trajectory for $N_\chi=80$ and $N_\chi=320$ stochastic
orbitals. The gaps were calculated by diagonalizing the KS Hamiltonian
for each configuration along the trajectory. The results are shown for
sGDM with $\gamma=0.5$ using ew-efsDFT. In the descent stage, the gap
changes markedly, while in the averaging stage, it fluctuates about an
average value, approaching the deterministic gap (black curve) as
$N_\chi$ increases. The fluctuations in the band gap result from
fluctuations in the structure and the electron density. The latter's
effect is summarized in Fig.~\ref{fig:gap}(b), where we plots the
standard deviation in the band gap for the equilibrium geometry as a
function of the number of stochastic orbitals. The standard deviation
of the band gap follows the expected $N_\chi^{-1/2}$, consistent with
the central limit theorem, with values on the order of several
meVs. We also find that sDFT always underestimates the gaps, as shown
in Fig.~\ref{fig:gap}(a) for two values of $N_\chi$. As shown in
Fig.~\ref{fig:gap}(d), the systematic error is slightly larger for the
optimized structures compared to sDFT calculation of the equilibrium
structure with a 5 meV difference. The systematic error scales
linearly as $N_\chi^{-1}$ which is consistent with previous
studies.\cite{doi:10.1002/wcms.1412}

Finally, we tested the stochastic optimization methods for Si$_{512}$
using ew-efsDFT. The results are shown in Fig.~\ref{fig:512}. Similar
to the case of Si$_{216}$ discussed above, BFGS provides the fastest
convergence of the RMSD, but it takes $\approx 30$ optimization steps
to achieve chemical accuracy compared $\approx 25$ optimization steps
for Si$_{216}$.  This slower convergence of the deterministic approach
is also observed for all stochastic optimization methods used in this
work. However, the conclusion drawn for Si$_{216}$ also holds for the
more extensive system; specifically, the optimal value suggested for
$\gamma$ in sGDM. In Fig.~\ref{fig:512}(b) we show a more direct
comparison of the optimization trajectories using sBFGS and sGDM for
two system sizes.  We note that the RMSD for both system sizes are
parallel in the descent stage for both stochastic optimization
methods, indicating that the optimization efficiencies of both sBFGS
and sGDM are comparable and are independent of the system
size. Fig.~\ref{fig:512}(c) shows reverse optimization trajectories
for the two system sizes, indicating that the accuracy in determining
the optimized structure is somewhat better for the larger systems,
likely due to self-averaging.  This effect, in fact, suggests that one
can reduce the number of stochastic orbitals used as the system size
increases and achieve sub-linear scaling for the same level of
accuracy.

\section{Conclusion}
\label{sec:con}
In this work, we assessed the efficiency and accuracy of obtaining the
ground state structure of extended systems by combining the linear
scaling sDFT to compute the forces on the nuclei with stochastic
optimization techniques, such as the stochastic gradient descent with
momentum and the stochastic BFGS approach. Typical optimization
trajectories can be divided into a descent step where the forces on
the nuclei are more significant than the fluctuations, followed by an
averaging stage.  We analyzed the role of noise, controlled by the
number of stochastic orbitals, the number of windows, and the size of
the fragments, in sDFT on the optimization trajectories for two
different system sizes.  We showed that both optimization methods
could efficiently determine the optimal structure of extended systems
with chemical accuracy by tuning the optimization parameters on small
systems.

\section*{Supplementary Material}
The asympototic behaviors of a reversed optimization trajectory with 
Langevin equation are provided in the supplementary material. 

\begin{acknowledgments}
We acknowledge support from the Center for Computational Study of
Excited State Phenomena in Energy Materials (C2SEPEM) at the Lawrence
Berkeley National Laboratory, which is funded by the U.S. Department
of Energy, Office of Science, Basic Energy Sciences, Materials
Sciences and Engineering Division under Contract No. DE-AC02-05CH11231
as part of the Computational Materials Sciences Program. Computational
resources were provided by the National Energy Research Scientific
Computing Center (NERSC), a U.S. Department of Energy Office of
Science User Facility operated under Contract
No. DE-AC02-05CH11231. R.B. gratefully acknowledges support from the
Germany-Israel Foundation (GIF) (Grant No. 201836). M.C. gratefully
acknowledge support from the Purdue startup funding.
\end{acknowledgments}

\section*{DATA AVAILABILITY}
All data that presented in this study are available from the corresponding 
author upon reasonable request.

\bibliographystyle{aipnum4-1}
\bibliography{stoopt}

\end{document}